\documentclass[%
 reprint,
superscriptaddress,
showkeys,
 amsmath,amssymb,
 aps,
pre, 
onecolumn
]{revtex4-2}

\usepackage{graphicx}% Include figure files
\usepackage[caption=false]{subfig} %?

\usepackage{dcolumn}% Align table columns on decimal point
\usepackage{bm}% bold math
\usepackage[%Uncomment any one of the following lines to test 
top=3cm, 
left=3cm,
right=3cm,
bottom=4cm,
]{geometry}

\newcommand{\tsr}{tSR} %abbreviation for standard/traditional SR
\newcommand{\dsr}{dSR} %abbreviation for dynamical SR
\newcommand{\xinit}{x_0} %abbreviation for initial spatial condition
\newcommand{\xtarget}{x_T} %abbreviation for the target position
\newcommand{\xreset}{x_r} %abbreviation for the resetting position

\newcommand{\ee}{\end{equation}} 
\newcommand{\be}{\begin{equation}}

\usepackage[mathscr]{euscript}  
\usepackage{xcolor}  

\usepackage{tcolorbox}
\usepackage{comment}
\usepackage{float}

\usepackage{hyperref}
\hypersetup{
    colorlinks=true,
    linkcolor=black,
    citecolor=black,
    filecolor=black,%magenta,      
    urlcolor=black,
    }

\makeatletter
\newsavebox{\@brx}
\newcommand{\llangle}[1][]{\savebox{\@brx}{\(\m@th{#1\langle}\)}%
  \mathopen{\copy\@brx\kern-0.5\wd\@brx\usebox{\@brx}}}
\newcommand{\rrangle}[1][]{\savebox{\@brx}{\(\m@th{#1\rangle}\)}%
  \mathclose{\copy\@brx\kern-0.5\wd\@brx\usebox{\@brx}}}
\makeatother

\usepackage{hyperref}
\hypersetup{
    colorlinks=true,
    linkcolor=blue,
    filecolor=magenta,      
    urlcolor=cyan,
    citecolor = magenta,
    pdftitle={Overleaf Example},
    pdfpagemode=FullScreen,
    }

\begin{document}

\preprint{ApS/123-QED}

\title{Optimizing Cost through Dynamic Stochastic Resetting} 

\author{Deepak Gupta}
\thanks{phydeepak.gupta@gmail.com}
\affiliation{UHasselt, Faculty of Sciences, Theory Lab, Agoralaan, 3590 Diepenbeek, Belgium}
\affiliation{Department of Physics, Indian Institute of Technology Indore, Khandwa Road, Simrol, Indore-453552, India}

\author{Bart Cleuren}
\thanks{bart.cleuren@uhasselt.be}
\affiliation{UHasselt, Faculty of Sciences, Theory Lab, Agoralaan, 3590 Diepenbeek, Belgium}

\begin{abstract}
The cost of stochastic resetting is considered within the context of a discrete random walk model. In addition to standard 
stochastic resetting, for which a reset occurs with a certain probability after \emph{each} step, we introduce a novel resetting protocol which we dubbed {\it dynamic resetting}. This protocol entails an additional dynamic constraint related to the direction of successive steps of the random walker. We study this novel protocol for a one-dimensional random walker on an infinite lattice. We analyze the impact of the constraint on the walker's mean-first passage time and the cost (fluctuations) of the resets as a function of distance of target from the resetting location. Further, cost optimized search strategies are discussed.
\end{abstract}

% \pacs{Valid pACS appear here} % pACS, the physics and Astronomy  Classification Scheme.
% \Keywords{Stochastic resetting, optimal search strategies}
\maketitle
%%%%%%%%%%%%%%%%%%%%%%%%%%

%\tableofcontents

%%%%%%%%%%%%%%%%%%%%%%%%%%%%%%%%%%%%%%%%%%%%%%%%%%%%%%%%%%%%%%%%%%
\section{Introduction}
Unraveling efficient search protocols is an attractive field of research in various scientific disciplines~\cite{search_mech-1, Sensing} ranging from biology~\cite{biology}, chemistry~\cite{chemistry}, to ecology~\cite{bell1991behavioural}. Animals foraging for food~\cite{foraging-1,foraging-2}, {\it E. coli} diffusing towards higher food concentration~\cite{berg2004coli}, and transcription factor finding DNA promoter sites~\cite{protein, Wang2013} are some of the examples of search mechanisms displayed by biological organisms. In computer science, researchers design efficient search algorithms to solve hard combinatorial problems~\cite{luby1993optimal,montanari2002optimizing}. Recent research discovered that sudden restart of a search process can dramatically change the diffusive searcher's mean first passage time to find the target~~\cite{evans2011diffusion}. In contrast, this mean first passage time of an unbiased random walker in the absence of restart protocol diverges. The common intuition behind how the restart mechanism expedites the search process is by truncating those trajectories that move further away from the target, and which would have contributed to longer search times. See Ref.~\cite{evans2020stochastic} and references therein for a detailed review of this topic.  

The traditional stochastic resetting (\tsr) mechanism is implemented by restarting/resetting the underlying (inherent) dynamical process at random time intervals~\cite{evans2011diffusion}. Thus, \tsr\ is an intermittent dynamical process whereby the overall mechanism involves slow excursions due to the inherent system's dynamics followed by sudden/instantaneous restarts. Several variants of \tsr\ have been explored in the past, notably including the space-~\cite{evans2011optimal} and time-dependent resetting~\cite{pal2016diffusion,shkilev2017continuous},  power-law resetting~\cite{nagar2016diffusion}, resetting in discrete-space and discrete-time models~\cite{DSDT-bound}, asymmetric resetting~\cite{20Plata}, and refractory period resetting~\cite{Bressloff2020intra, Evans_2019, Valladares_2024}. Several applications of stochastic resetting have been demonstrated for a wide variety of processes, such as the Mpemba effect~\cite{21Busiello,bao2022accelerating}, molecular dynamics simulations~\cite{Blumer2024}, overfitting protocols~\cite{overfitting}, speed-limit~\cite{SL-UD}, fast equilibration techniques~\cite{goerlich2024resetting}, erasure~\cite{goerlich2023experimental}, Maxwell's demon~\cite{Demon}, and income dynamics~\cite{Santra_2022,arnab-income}. 

In contrast to the \tsr\ protocol, in this paper, we introduce a novel {\it dynamic stochastic resetting} (\dsr) protocol. This protocol is inspired by, and related to, a resetting mechanism previously considered in the context of Brownian Donkeys~\cite{Bart-C-paper}. These donkeys are a type of Brownian motors with absolute negative mobility, meaning the peculiar property that the motor on average moves against an externally applied bias (see for example~\cite{Eichhorn_anm_PRL2002,Cleuren_anm_PRE2003, Ros_anm_NAT2005,MachuraPRL2007,Sarracino_anm_PRL2016} for other representative examples). In the \dsr\ protocol, a random walker (depicting the underlying dynamical process) is allowed to reset (to a predetermined location) only when the \emph{number of consecutive steps in the same direction} (later defined as $\sigma$) reaches a threshold value $N_r$. The threshold value $N_r$ hence represents the minimal number of consecutive steps in the same direction before a reset can occur. Upon a change in direction or a reset, the value of $\sigma$ is re-initialised. In the limit of $N_r=1$, both frameworks (\dsr\ and \tsr) are identical. We stress that \dsr\ is different than the space-dependent resetting framework~\cite{evans2011optimal, 20Plata}, where in the latter the resetting occurs when the process crosses a threshold value. 

In this paper, we apply the \dsr\ protocol on a random walker on a one-dimensional infinite lattice. Herein, we compute the mean first passage time (MFPT) and its standard deviation, and the coefficient of variation, each as a function of the threshold value $N_r$ and the target location. Then, we extend our analysis to estimate the cost of dynamical stochastic resetting for an ensemble of the first passage trajectories. We highlight that thermodynamic cost of stochastic resetting has been previously investigated for different settings including instantaneous resetting~\cite{Deepak_PRL, Morientropy}, proportional~\cite{Olsen_2024-PR} and stochastic return~\cite{Deepak2022_work, olsen2023thermodynamic}, first passage process~\cite{Sunil_2023, Pal_tradeoff}, unidirectional process~\cite{Deepak_PRR,fuchs2016stochastic}, uncertainty relations~\cite{uncertainty-pal}, and intermittent switching potentials~\cite{goerlich2023experimental, goerlich2024resetting}.

We investigate the impact of \dsr\ on the first passage time and the cost of resetting and compare these results with \tsr. We motivate ourselves from the perspective of computer search algorithms, where one requires time-efficient protocols to solve hard combinatorial problems~\cite{luby1993optimal,montanari2002optimizing}. \tsr\ is an interesting route to reduce effective computational time~\cite{FPT-pal}. However, one of the shortcomings of \tsr\ is that the MFPT diverges for stronger resetting (in the limit of resetting's frequency going to infinity). On the contrary, we will show that \dsr\ reduces the MFPT (in comparison to \tsr) in this particular limit. We also discuss the cost-effectiveness of \dsr's search protocols. To the best of our knowledge, such a dynamic resetting protocol was not investigated earlier, which also motivates us to study herein. 

This paper is organized as follows. Section~\ref{setup} describes the algorithm of the dynamical stochastic resetting protocol for a one-dimensional random walk. Section~\ref{result} discusses the statistics of the first passage time and the mean cost of dynamic resetting. We summarize our paper in Sec.~\ref{discussion}.  The stationary probability distribution of random walker for \tsr\ is discussed in Appendix~\ref{sec:pdf}. The calculations for the moment generating functions of the first passage time and cost for \tsr\ are, respectively, relegated to Appendices~\ref{sec:mgf-fpt} and \ref{sec:mgf-cost}. We discuss the mean first passage time in the limit of the resetting probability $p_r\to 1$ in Appendix~~\ref{sec:mfpt-pr-to-1}. Appendix~\ref{sec:window-resetting} presents a comparison of the mean first passage time obtained using the window resetting (discrete analogy of Ref.~\cite{evans2011optimal}) with that of \dsr. We discuss the mean cost for the case when the threshold distance is larger than the target distance in Appendix~\ref{sec:cost-nr-gr-xt}. Method of \dsr\ simulations is discussed in Appendix~\ref{methods}.

\section{Setup}\label{setup}
We consider a random walker (RW) on a one-dimensional (1D) infinite lattice. The spatial position is labeled by $x\in\mathbb{Z}$. In the context of resetting, the RW starts at an initial position $\xinit$ and moves about until it hits the target at position $\xtarget$. Additionally, we define $\xreset$ as the spatial position towards the RW is relocated upon a resetting event. Jumps to the right (left) occur with probability $p~(q)$ such that $p+q = 1$. Hence, both space and time are discrete. For the resetting mechanism we introduce an additional counter $\sigma$ which keeps track of the length of the last sequence of successive steps taken by the RW \emph{in the same direction}. This counter can be 0 (at the initial condition and also immediately after a reset occurs) or positive/negative if the last step was to the right/left. A step in the same direction as the previous one will increase (for consecutive steps to the right) the counter by $+1$, and decrease the (negative) counter by $-1$ for consecutive steps to the left. A change in direction of the RW sets the counter to $+1$ ($-1$) if previous steps were to the left (right) followed by a step to the right (left). In the dynamical protocol a reset is only allowed, with a certain probability $p_r$, when this counter is greater or equal to a threshold value, that is when $\vert \sigma \vert \geq N_r$. Hence, as stated before, this parameter $N_r$ represents the minimal number of consecutive steps in the same direction before a reset can occur. Clearly, when $N_r =1$ we recover standard stochastic resetting tSR.

\begin{figure}
    \centering
    \includegraphics[width = \textwidth]{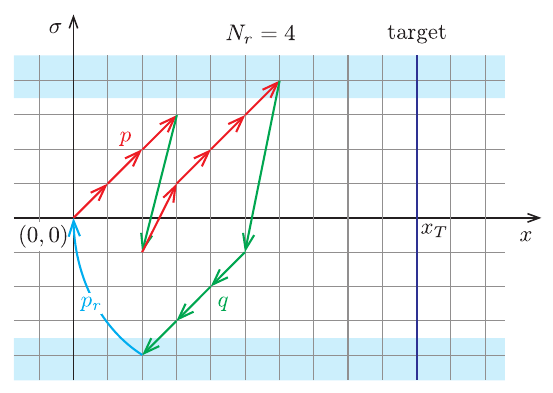}
    \caption{Sketch of a trajectory, starting at location $(0,0)$, made by a random walker (RW) on a 1D infinite lattice $x\in\mathbb{Z}$, where $x$ denotes the RW's position. The counter, $\sigma$, counts the number of steps made by RW in one direction. Hops to the right (left) occur with probability $p~(q=1-p)$. Resets to the resetting location $x=0$ occur with a probability $p_r$ whenever $|\sigma| \geq N_r$ (indicated by the blue regions). The counter, $\sigma$, is set to $-1~(+1)$ when the RW switches its spatial direction from positive to negative (negative to positive). The solid blue line, at position $\xtarget$ marks the position of the target: if the RW is on any point along this line after a complete move is made, the target is found and the movement stops.}
    \label{fig:scheme}
\end{figure}

It is clear that $x$ alone is no longer a Markovian variable, and one needs to consider both $x$ and $\sigma$ in order to describe the full dynamics. As the RW moves about, a trajectory in the $(x,\sigma)$-plane is traced out. An example is given in Fig.~\ref{fig:scheme}: a RW starts at the origin $(0,0)$ and explores the space until a reset occurs at position $(2,-4)$ after which the RW ends up back in the origin $(0,0)$. As previously described, and as is made evident by the figure, as long as the RW moves in \emph{the same} direction, the counter $\sigma$ is increased (for a spatial jump to the right) or decreased (for a spatial jump to the left) by $1$. A \emph{change of direction} sets the counter to either $+1$ when the last jump was to the right and the jump(s) before the last one to the left, or $-1$ in the opposite case.
A \emph{complete move} or \emph{time step} in the $(x,\sigma)$-plane involves two steps: first a spatial jump is made, as in the classical random walk setting. As a result of this jump both $x$ and $\sigma$ are updated. As a second step, given the updated value of $\sigma$, a check is made to determine whether $\sigma$ reaches the threshold, that is whether $N_r \leq \vert \sigma \vert$. If this is the case, with probability $p_r$ a reset is done. So both steps (first the spatial jump, then the check for a reset) comprises one complete move in the $(x,\sigma)$-plane. This is repeated until, after a complete move is done, the RW has reached the target located at a given position $\xtarget$. Note that the complete move implies that even if the RW is at $\xtarget$ immediately after the spatial jump, one still has to check for a reset (which can bring the RW back to the origin). If a reset occurs, the target in fact is not yet reached. The diagram below gives all possible moves, starting from the current state $(x,\sigma)$, together with the probability of the move and the conditions on $\sigma$:
\begin{align}\label{eqn:fdrp}
    (x,\sigma) \xrightarrow{} 
    \begin{cases}
    (x+1,\sigma + 1)&\quad{\rm with~probability~} p~{\rm for}~0 \leq \sigma < N_r-1 \\
    (x+1, \sigma + 1) &\quad {\rm with~probability~} p(1-p_r)~{\rm for}~N_r-1 \leq \sigma \\
    (0,0) &\quad {\rm with~probability~} pp_r~{\rm for}~N_r-1 \leq \sigma\\
    (x-1, \sigma-1) &\quad {\rm with~probability~} q~{\rm for}~-N_r+1 < \sigma \leq 0 \\
    (x-1,\sigma-1)&\quad {\rm with~probability~} q(1-p_r)~{\rm for}~\sigma \leq -N_r+1 \\
    (0,0) &\quad {\rm with~probability~} qp_r~{\rm for}~\sigma \leq -N_r+1\\
         (x + 1, 1)&\quad {\rm with~probability~} p ~{\rm for}~\sigma < 0\\
     (x - 1,-1)&\quad {\rm with~probability~} q ~{\rm for}~\sigma > 0\ ,
    \end{cases}
\end{align}

\begin{figure}[t!]
    \centering
    \includegraphics[width = \textwidth]{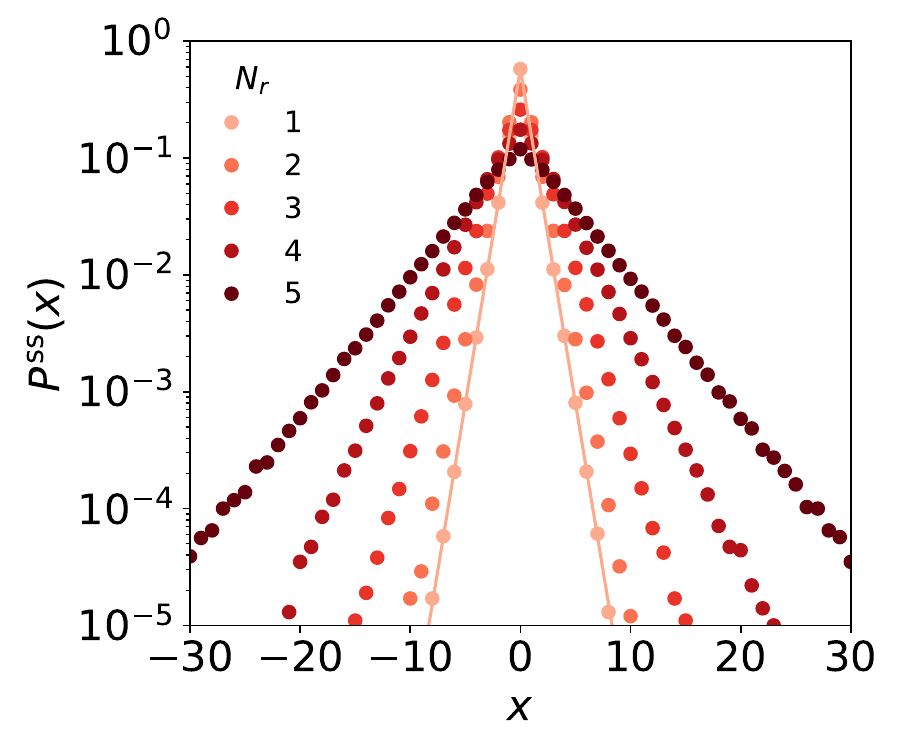}
    \caption{Unbiased random walk ($p=q=1/2$). Stationary probability distribution $P^{\rm ss}(x)$ as a function of position $x$. Color intensity increases with $N_r$. Resetting probability $p_r = 0.5$. Solid line: Analytical prediction for $N_r=1$~\eqref{pss-x}. Time steps: $10^3$. Number of realizations: $10^6$. }
    \label{fig:pdf}
\end{figure}

In the absence of a target, the position distribution of the RW will reach a non equilibrium stationary state. Figure~\ref{fig:pdf} shows the comparison between the stationary distributions (numerical simulation data) obtained for different threshold values $N_r$. We also show the comparison between analytical prediction~\eqref{pss-x} for $N_r=1$ with the numerical simulation data. As is intuitively clear, the distribution becomes wider as $N_r$ increases. This is expected as the likelihood for a reset decreases for increasing $N_r$.

In this paper, our first objective is to study the statistics of the first passage time for different values of threshold distance $N_r$, 
as a function of target's distance, $\xtarget$, from RW's resetting location. The first passage time is the time taken by the RW starting from $x(0)$ and hitting the target at $\xtarget$ for the first time:
\begin{align}\label{fpt}
    n_{\rm FP} \equiv {\rm inf}\{n ; x(n) = \xtarget|x(0)\} \ .
\end{align}
The second quantity of the interest is the cost~\cite{Sunil_2023} of resetting until the random walker hits the target $\xtarget$ for the first time:
\begin{align}\label{cost-0}
    C_\beta\equiv c\sum_{i = 1}^{\mathcal{N}(n_{\rm FP})} |x(i) - \xreset|^\beta\ ,
\end{align}
where $c$ is the intrinsic cost associated with each resetting. $x(i)$ and $\xreset$, respectively, are RW's position before and after the $i$-th resetting, $\beta$ is an exponent, and $\mathcal{N}(n_{\rm FP})$ is the number of reset events occurred before the system hitting the target for the first time. For convenience, we rescale $C_\beta$ by $c$, such that the former becomes a dimensionless quantity:
\begin{align}\label{cost}
    C_\beta = \sum_{i = 1}^{\mathcal{N}(n_{\rm FP})} |x(i) - \xreset|^\beta\ .
\end{align}
Both first passage time $n_{\rm FP}$~\eqref{fpt} and cost of reset $C_\beta$~\eqref{cost} are stochastic quantities. And while analytical computation of their statistics is untractable for $N_r > 1$, for the case $N_r=1$ one can compute analytically the exact moment generating function of first passage time and the cost $C_\beta$. These are respectively given below (see Appendices~\ref{sec:mgf-fpt} and \ref{sec:mgf-cost} for detailed calculations):
\begin{align}
\tilde{F}(z|\xinit)& = \dfrac{\tilde{F}_{\rm NR}(z(1-p_r)|\xinit)}{1-zp_r \tilde{S}_{\rm NR}(z(1-p_r)|\xinit) }\ ,\label{fpt-st-main}\\ 
\tilde{\bar{\Phi}}(k,1|\xinit)& = \dfrac{\tilde{F}_{\rm NR}(1-p_r|\xinit)}{1-p_r \tilde{\bar{\phi}}_{\rm NR}(k, 1-p_r|\xinit) }\ .\label{mgf-cost-main}
\end{align}
For simplicity, here and in what follows, we consider the initial location $\xinit$ and resetting location $x_r$ to be the same, i.e.,  $\xreset= \xinit$.
The quantities $\tilde{S}_{\rm NR}(z|\xinit)$ and $\tilde{F}_{\rm NR}(z|\xinit)$, respectively, are the $z$-transformed~\footnote{We define the $z$-transform as $\tilde{g}(z) \equiv \sum_{n=0}^{\infty}g(n)z^n$.} survival probability and first passage distribution of the non-reset (NR) random walker starting from $x= \xinit$ in the presence of an absorbing boundary at $x = \xtarget$.  $\tilde{{\phi}}_{\rm NR}(k, z|\xinit)$ is the $z$-transformed reset-free moment generating function of the cost, averaged over ensemble of trajectories in the presence of absorbing boundary at $\xtarget$:
\begin{align}
    \tilde{{\phi}}_{\rm NR}(k, z|\xinit)\equiv \sum_{x=-\infty}^{\xtarget}~e^{ik|x-\xinit|^\beta}\tilde{P}^{\rm abs}_{\rm NR}(x,z|\xinit)\ .
\end{align}
Here, $\tilde{P}^{\rm abs}_{\rm NR}(x,z|\xinit)$ is the $z$-transformed probability distribution function of random walker's positions, starting from $x=\xinit$ and in the presence of absorbing boundary at $x=\xtarget$. Using Eqs.~\eqref{fpt-st-main} and ~\eqref{mgf-cost-main}, we can compute the moments of first passage time and cost, and these are shown in Appendices~\ref{sec:mgf-fpt} and \ref{sec:mgf-cost}.

For threshold distances $N_r>1$ we compute the first passage time and cost using \dsr\ discussed in Eq.~\eqref{eqn:fdrp} by using the numerical computations (see Appendix~\ref{methods} for the methods of simulations).

\section{Results}\label{result}
In what follows we focus, for simplicity, on the symmetric random walk $p=q=1/2$. First, we discuss the statistics of the first passage time, and then, show the results for the mean cost of resetting.  

\begin{figure}[t!] % FIG 3
    \centering
    \includegraphics[width = \textwidth]{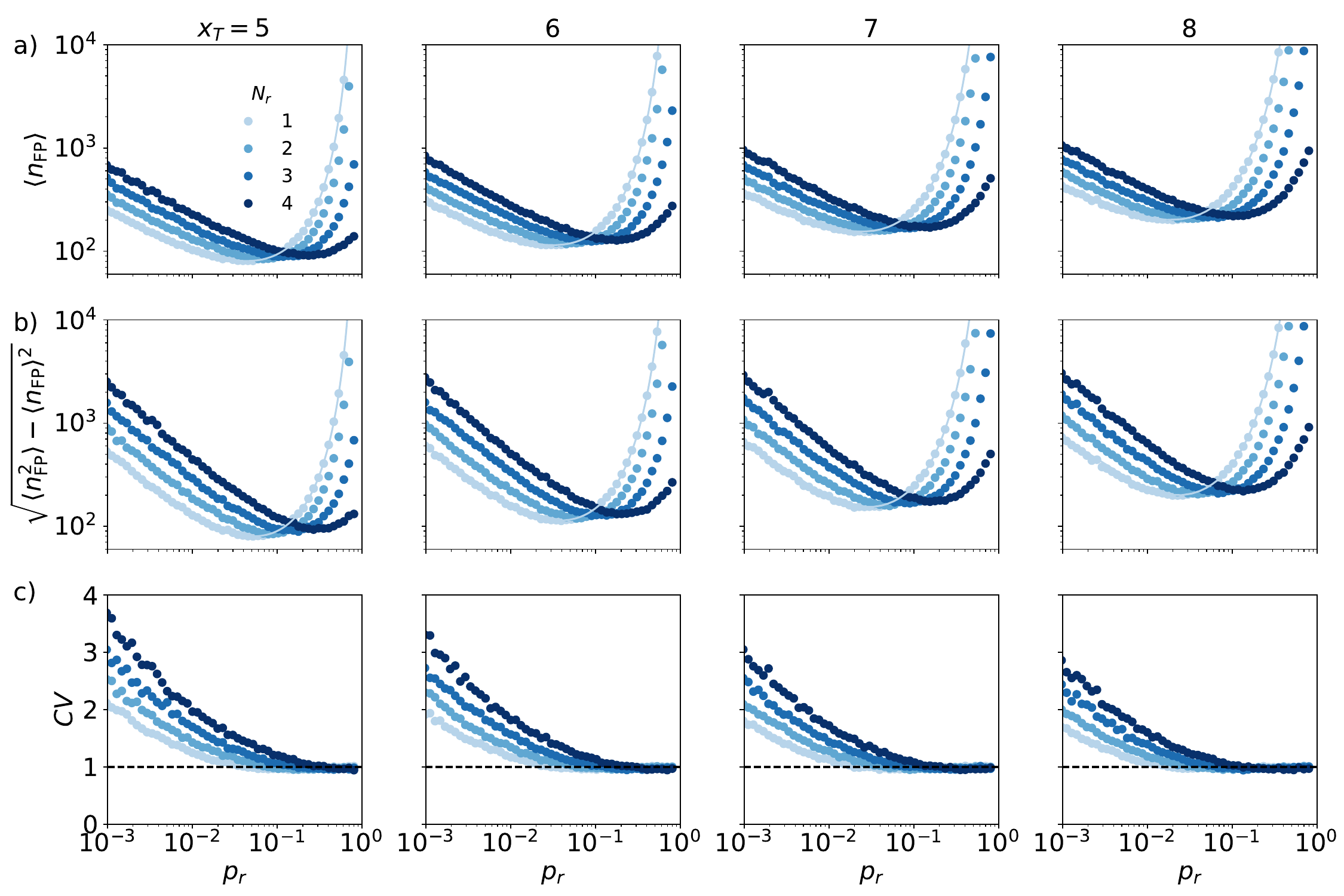}
    \caption{Unbiased random walk $(p=q=1/2)$. (a) Mean first passage time $\langle n_{\rm FP} \rangle$, (b) standard deviation $\sqrt{\langle n_{\rm FP}^2 \rangle - \langle n_{\rm FP} \rangle^2}$, and (c) coefficient of variation $CV\equiv \dfrac{\sqrt{\langle n_{\rm FP}^2 \rangle - \langle n_{\rm FP} \rangle^2}}{\langle n_{\rm FP} \rangle}$, each as a function of reset probability $p_r$. $\xtarget$: Target's distance from the resetting location. The color intensity increases with the threshold distance $N_r$. (a-b) Solid curves are the analytical predictions [Eqs.~\eqref{mfpt-exp} and \eqref{var-exp}] for $N_r=1$. (c) Horizontal dashed line corresponds to $CV=1$. For all plots, resetting $\xreset$ and initial location $\xinit$ are same (i.e., $\xreset = \xinit$) and number of realizations $10^4$.}
    \label{fig:mfpt}
\end{figure}

In the following, we discuss the first passage time for the case when the threshold distance is smaller than the target distance from the resetting location, i.e., $1\leq N_r < \xtarget$. Figure~\ref{fig:mfpt}a discusses the mean first passage time (MFPT), $\langle n_{\rm FP} \rangle$, as a function of the resetting probability $p_r$ for different threshold $N_r$ and target distances $\xtarget$. Moreover, for $N_r=1$, we show the comparison of numerical simulation result with the analytical result~\eqref{mfpt-exp}. For each case, the MFPT diverges in the limit $p_r\to 0$. 
\begin{figure}[h!] % FIG 4
    \centering
    \includegraphics[width = \textwidth]{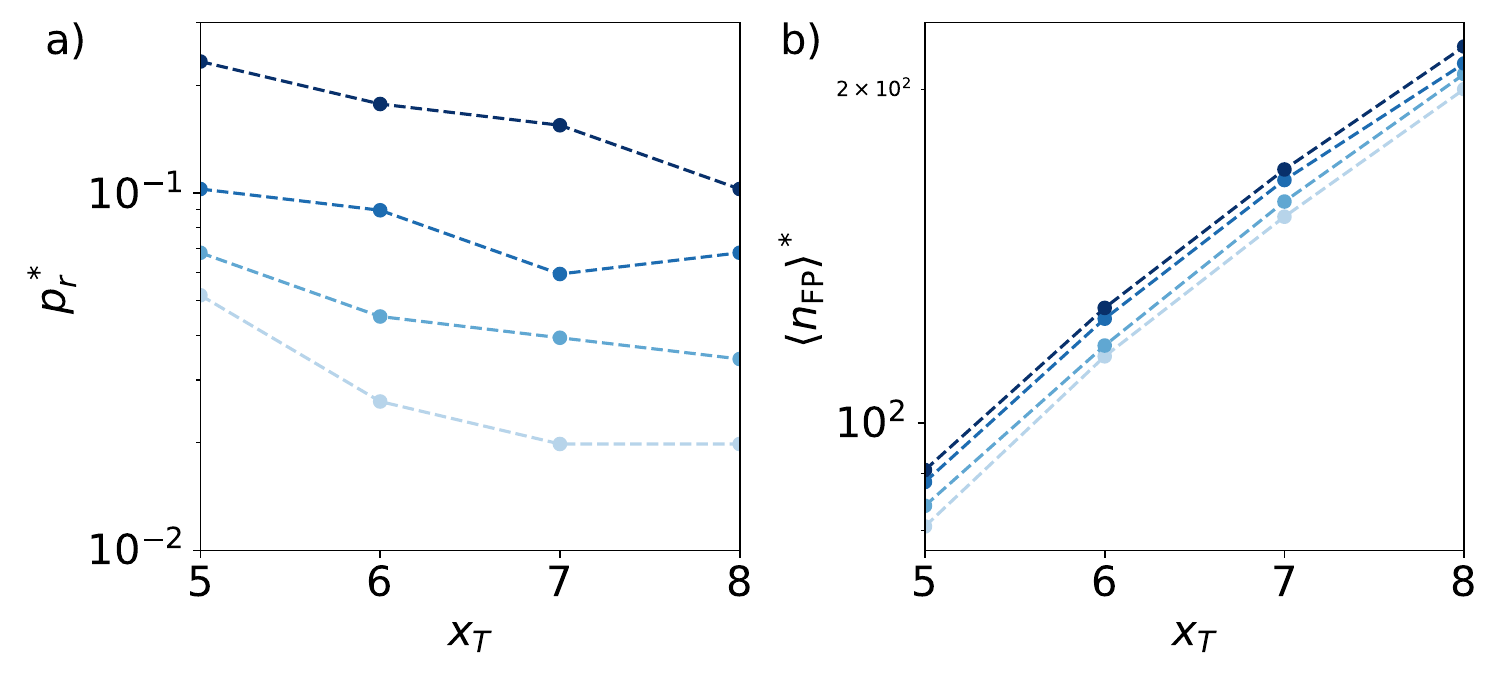}
    \caption{Unbiased random walk $(p=q=1/2)$. (a) Optimal resetting probability $p_r^*$ and (b) optimal mean first passage time $\langle n_{\rm FP}\rangle^*$, each as a function of target's distance from the resetting location $\xtarget$. The color intensity increases with threshold distance $N_r$ (Fig.~\ref{fig:mfpt}). Connecting lines are guide to the eye. }
    \label{fig:optimal-mfpt}
\end{figure}
\begin{figure}[b!]% FIG 5
    \centering
    \includegraphics[width = \textwidth]{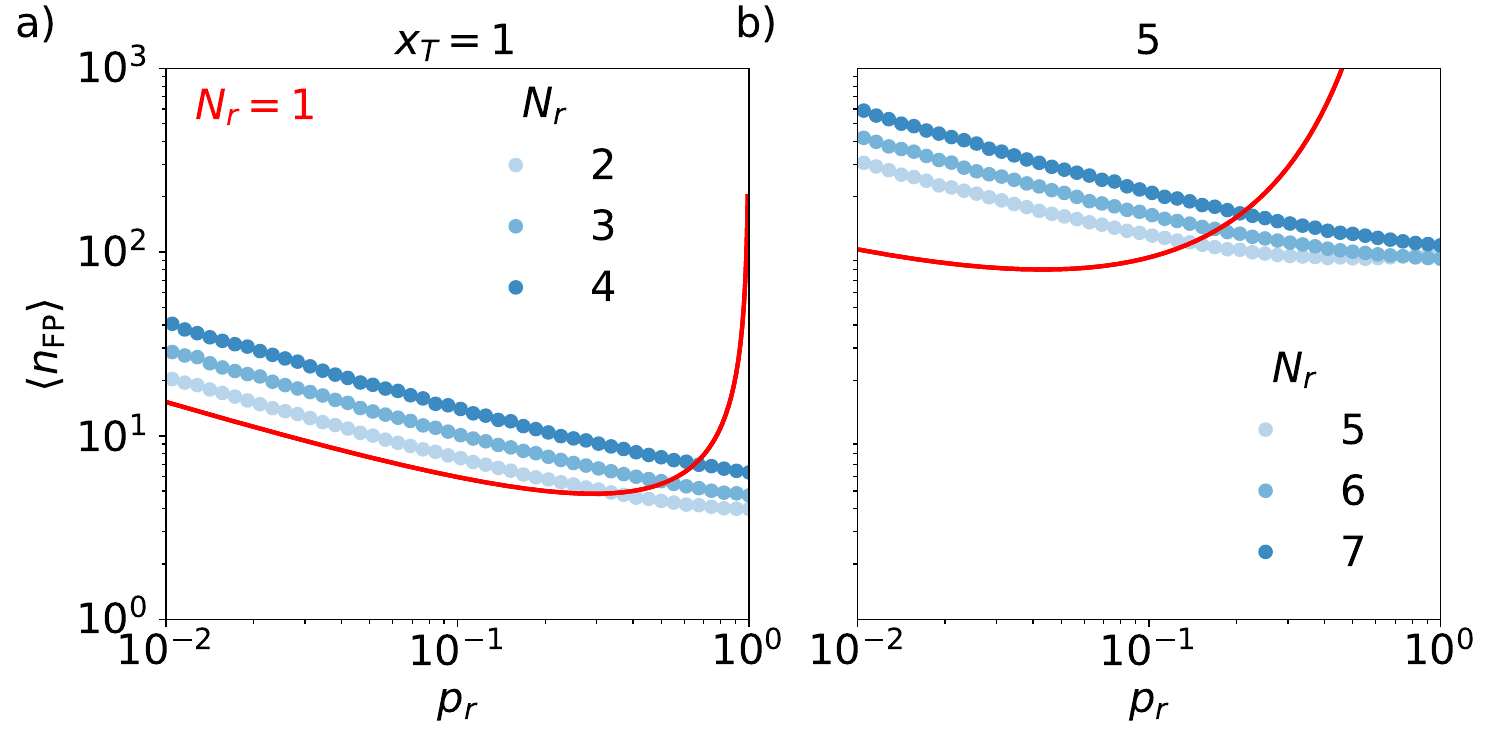}
    \caption{Unbiased random walk ($p=q=1/2$). Mean first passage time, $\langle n_{\rm FP}\rangle$, as a function of resetting probability $p_r$. Symbols: Numerical simulation data. Red curve: Analytical result for $N_r=1$. Blue: a) $\xtarget = 1$ and $N_r = 2,3,4$, and b) $\xtarget = 5$ and $N_r = 5,6,7$.}
    \label{fig:mfpt-Nr-gth-xb}
\end{figure}
\begin{figure}% FIG 6
    \centering
    \includegraphics[width = \textwidth]{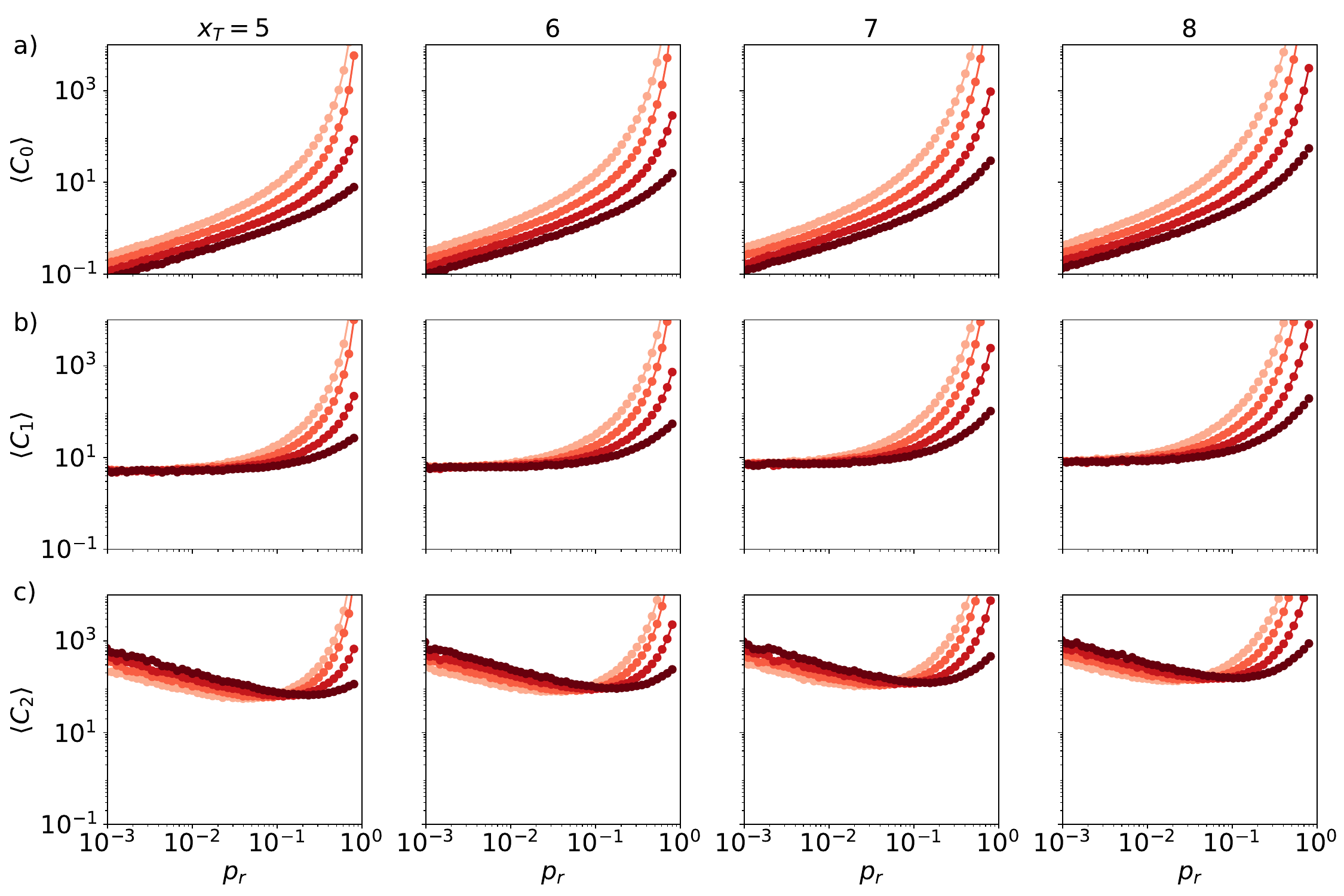}
    \caption{Unbiased random walk ($p=q=1/2$). Mean cost~\eqref{cost} as a function of reset probability $p_r$. The color intensity increases with threshold distance $N_r$ (Fig.~\ref{fig:mfpt}). Connecting lines are guide to the eye.}
    \label{fig:mfpt-cost-comb}
\end{figure}
\begin{figure}[b!]% FIG 7
    \centering
    \includegraphics[width = \textwidth]{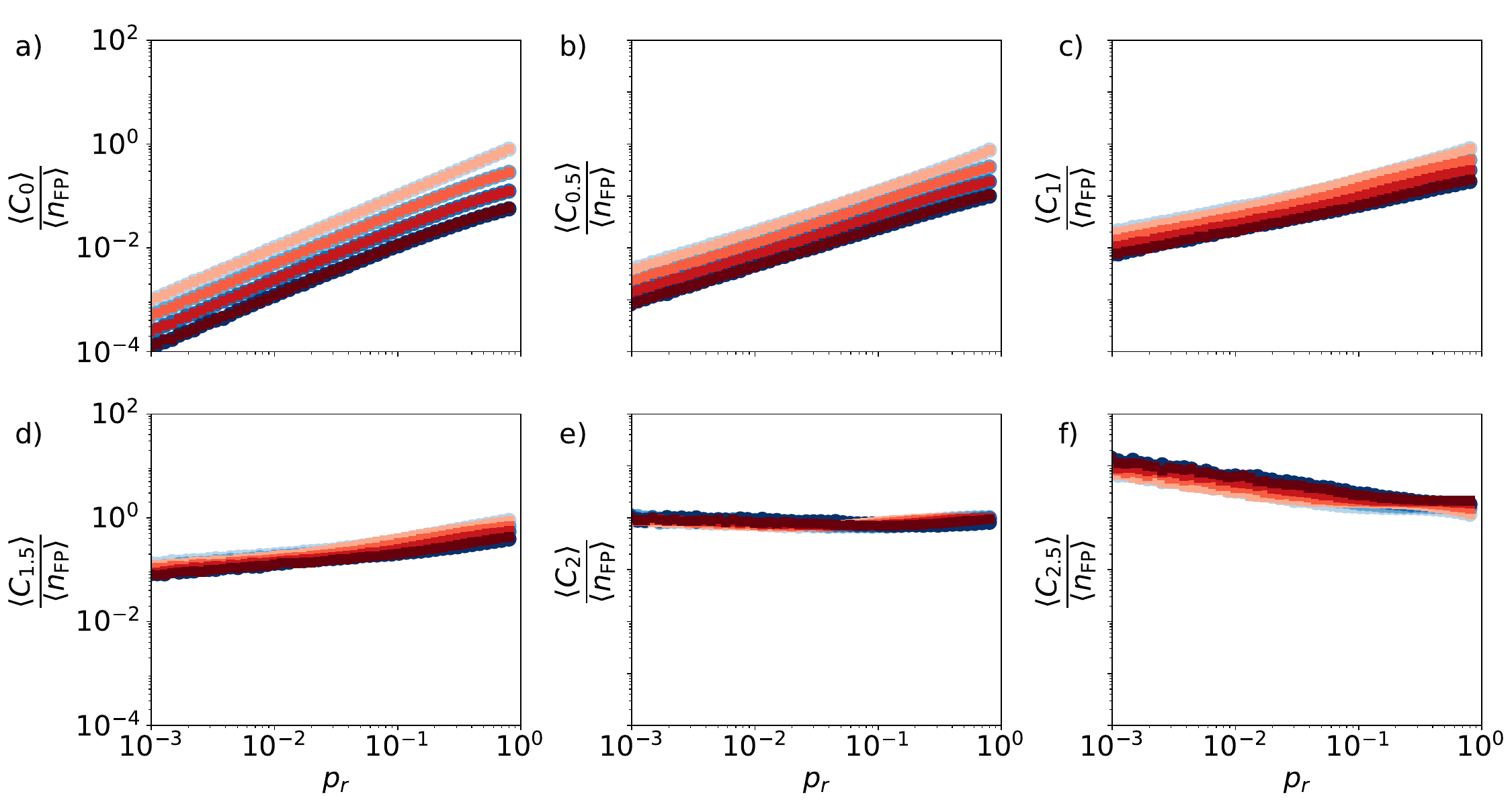}
    \caption{Unbiased random walk ($p=q=1/2$). Ratio of average cost, $\langle C_\beta\rangle$~\eqref{cost} to the mean first passage time $\langle n_{\rm FP} \rangle$, as a function of resetting probability $p_r$. The red (blue) color intensity increases with threshold distance $N_r:1-4$ (target's distances $\xtarget = 5,8$). For each red color (fixed $N_r$), (blue) data is collapsed for two target's distances $\xtarget = 5,8$.}
    \label{fig:cost0}
\end{figure}
This is to be expected, as this limit  corresponds to a reset-free random walk for which the MFPT diverges. In the opposite limit ($p_r\to 1$), the MFPT diverges for $N_r=1$ and $2$, respectively, for all $|\xtarget| > 0$ and $|\xtarget| > 1$ (Fig.~\ref{fig:mfpt-pr-1}). This is because for $N_r=1$ the walker (starting from origin) forever stays at the origin (i.e., the resetting location), and for $N_r=2$ the walker only explores the lattice points $-1,0,1$. For larger $N_r > 2$ the MFPT is finite for $p_r= 1$ (Fig.~\ref{fig:mfpt-pr-1}). It is clear form Figure~\ref{fig:mfpt}a that the MFPT has a global minimum for each $N_r$. For a fixed $p_r \ll p^*_{r}(N_r=1)$, the MFPT increases with $N_r$, where $p^*_{r}(N_r)$ is the optimal resetting probability for $N_r$:
\begin{align}
    p_r^*(N_r) \equiv \underset{p_r \in [0,1]}{\rm argmin} \langle n_{\rm FP} \rangle\ . \label{eqn:opt-pr}
\end{align}
This is because the dynamical resetting protocol reduces the probability of a reset as compared to the standard resetting protocol with $N_r=1$. Hence, the MFPT increases by those trajectories which wander off in the opposite direction of the target. On the other hand, for $p_r \gg p^*_{r}(N_r=1)$, a larger $N_r$ allows the RW to explore more space before a reset occurs. And this helps the random walker to find the target. Increase in the target's distance increases the MFPT for each fixed $N_r$, as expected. Standard deviation (Fig.~\ref{fig:mfpt}b) of the first passage time also displays similar behavior. 
Figure~\ref{fig:mfpt}c shows the coefficient of variation $CV\equiv \frac{\sqrt{\langle n_{\rm FP}^2 \rangle - \langle n_{\rm FP} \rangle^2}}{\langle n_{\rm FP} \rangle}$ as a function of $p_r$. In the limit $p_r\to 0$, the order of fluctuation relative to mean goes higher for higher threshold distance $N_r$; this implies the first passage distribution becomes wider as $N_r$ increases. Surprisingly, for larger resetting probability, $CV\approx 1$ for all cases.

Figure~\ref{fig:optimal-mfpt}a discusses the variation of the optimal resetting probability $p_r^*$~\eqref{eqn:opt-pr} as a function of the target's distance, $\xtarget$. For each $N_r$, $p_r^*$ decays as $\xtarget$ increases. This is expected because one has to reduce the resetting probability so that the random walker could explore more space in order to hit the target placed far away. Eventually, this increases the optimal MFPT $\langle n_{\rm FP} \rangle_{p_r\to p_r^*}$ (Fig.~\ref{fig:optimal-mfpt}b). Moreover, Fig.~\ref{fig:optimal-mfpt}b shows that MFPT has a global minimum with respect to $N_r$ at $N_r=1$.

Figure~\ref{fig:mfpt-Nr-gth-xb} discusses the case for $N_r \geq \xtarget$, and it shows that for $N_r > \xtarget$, optimal MFPT is achieved for $p_r\to 1$. This is because \dsr\ resets a {\it specific} fraction of trajectories which perform directed random walk in consecutive steps until it hits the threshold $N_r$, other fraction of trajectories will hit the absorbing boundary without experiencing resetting. (This is in contrary to \tsr\ where each trajectory can be reset with probability $p_r$.) Moreover, the probability of resetting these trajectories increases with $p_r$; therefore, the mean first passage time decreases monotonically. For each fixed target location, MFPT is optimized with respect to $N_r$, and its optimal value is $N_r^* = \xtarget + 1$ (see also Fig.~\ref{fig:mfpt-pr-1} for $p_r=1$). Additionally, for $N_r \to \infty$, MFPT is expected to diverge for all $p_r$, since this limit corresponds to a reset-free random walker. Finally, in Fig.~\ref{fig:WSR-DSR} and Appendix~\ref{sec:window-resetting}, we discuss a comparison of the mean first passage time between the dynamic and window stochastic resetting (in analogy with Ref.~\cite{evans2011optimal}). It shows that dynamic stochastic resetting is better in the limit $p_r\to 1$ when $N_r<\xtarget$; otherwise, for $N_r>\xtarget$ window resetting has lower mean first passage time for all $p_r$.

Next we turn our attention towards the cost of resetting $C_\beta$ for $N_r<\xtarget$. Figure~\ref{fig:mfpt-cost-comb} shows mean cost of resetting, $\langle C_{\beta} \rangle$~\eqref{cost}, as a function of resetting probability, $p_r$, for different threshold $N_r$ and target distances, $\xtarget$. (See Fig.~\ref{fig:analytical-sim-cost} for the comparison between analytical and numerical simulation results for $N_r=1$.) In each panel of Fig.~\ref{fig:mfpt-cost-comb}a, the mean constant cost $\langle C_{0} \rangle$ (or the mean number of resets)  increases monotonically. This is expected since increasing $p_r$, number of resets also increases. For each fixed $p_r$, $\langle C_{0} \rangle$ reduces as $N_r$ increases. This is because increasing $N_r$ reduces the probability of resetting. $\langle C_{0} \rangle$ increases as $\xtarget$ increases, as expected. Linear cost $\langle C_{1} \rangle$ also increases with $p_r$ (Fig.~\ref{fig:mfpt-cost-comb}b), whereas the quadratic cost $\langle C_{2} \rangle$ shows non-monotonic behavior. In the limit $p_r\to 0$, $\langle C_{1,2} \rangle \neq 0$. This is because herein whenever the reset happens, the walker returns from very far distance, this gives non-zero value of the mean cost~\cite{olsen2023thermodynamic, Sunil_2023}. The behavior of the mean cost strongly depends on the exponent $\beta$~\eqref{cost}. Additionally, we find for $N_r\geq \xtarget$ results are qualitatively similar (Fig.~\ref{fig:cost-nr-gr-xt}).

Figure~\ref{fig:cost0} shows the ratio of mean of constant cost $\langle C_{\beta} \rangle$ to MFPT as a function of resetting probability $p_r$. In each panel, for each $N_r$, the data is plotted for two target's distances $\xtarget =5,8$. Figure~\ref{fig:cost0}(a-d) shows this ratio is independent of target distance, whereas we observe weak dependence on target distance in Fig.~\ref{fig:cost0}(e-f). $\langle C_{0} \rangle/\langle n_{\rm FP} \rangle = p_r$ [Eq.~\eqref{c0-formula}] for $N_r=1$. This is expected because MFPT increases with the target distance so does the mean number of resets; this makes the ratio to be independent of $\xtarget$. For $\beta=0-1.5$, mean cost per MFPT reduces with increasing $N_r$, and for $\beta=2$, this ratio insensitive to the value of $N_r$, where this ratio increases with $N_r$ for $\beta = 2.5$.

\begin{figure}
    \centering
    \includegraphics[width = \textwidth]{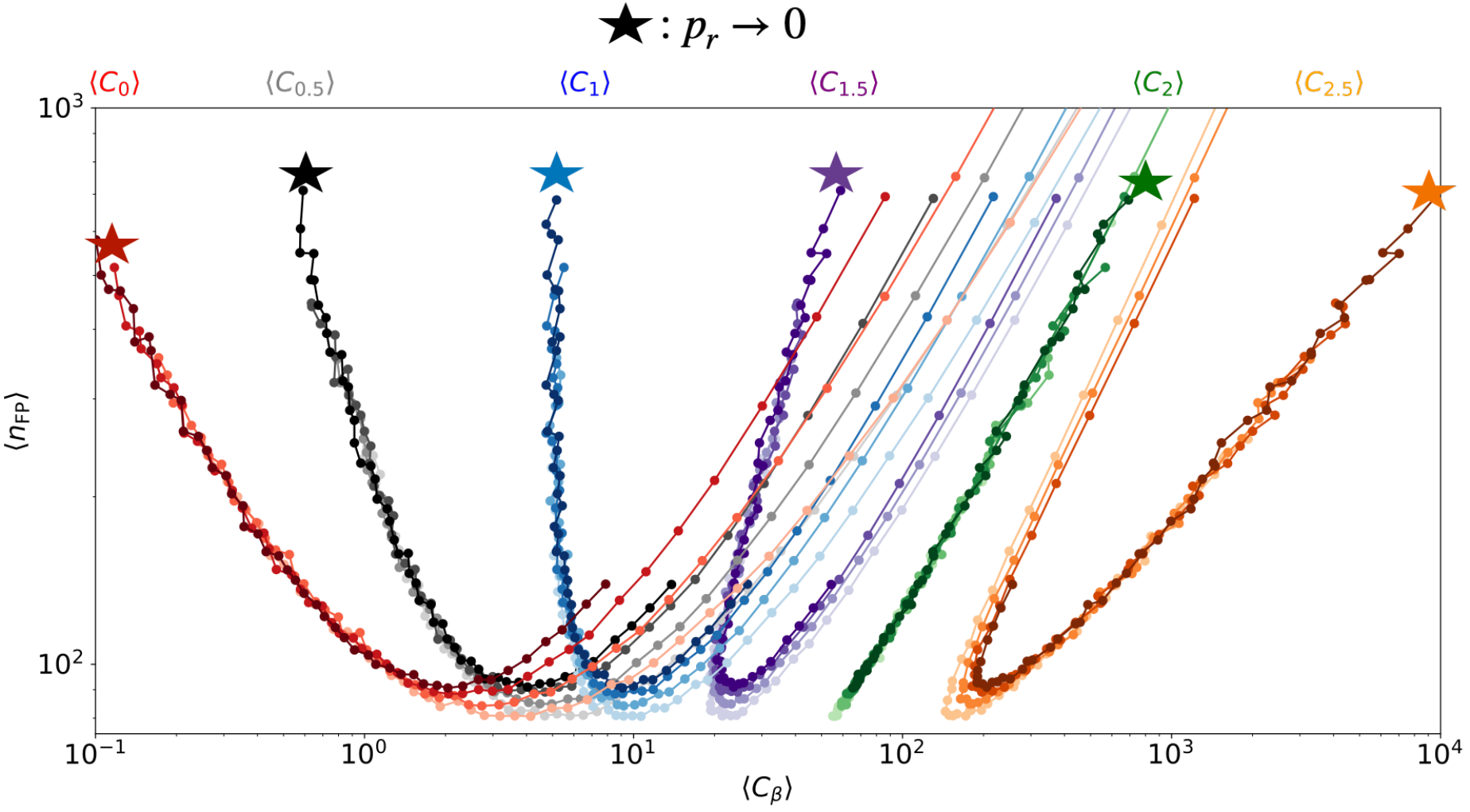}
    \caption{Unbiased random walk ($p=q=1/2$). Mean first passage time $\langle n_{\rm FP}\rangle$ as a function of mean cost $\langle C_{\beta}\rangle$~\eqref{cost} for target distance $\xtarget=5$. The color intensity increases with threshold distance $N_r$ (Fig.~\ref{fig:mfpt}). Connecting lines are guide to the eye. Star indicates the direction of $p_r\to 0$ limit.}
    \label{fig:mfpt-and-cost}
\end{figure}

Figure~\ref{fig:mfpt-and-cost} discuss the trade-off relation between the mean first passage time and the mean cost of resetting $\langle C_{\beta}\rangle$ for different exponents $\beta$. Each curve is parameterized by the resetting probability, $p_r$. In the limit of $p_r\to 0$, the behavior is independent of $N_r$. (Star symbol indicates the direction of $p_r\to 0$ limit.) For each exponent $\beta$ and $N_r$, MFPT can be optimized with respect to the mean cost, and the former has a global minimum for $N_r=1$ for all cases (see the lightest colored symbols). For $\beta<2$, the cost to achieve a certain MFPT is lower for larger $N_r$, whereas the trends is opposite for $\beta>2$ in the limit $p_r \to 1$.  

\section{Conclusion}\label{discussion}
In this paper, we investigated the dynamic stochastic resetting protocol, whereby a system resets with probability $p_r$ only when it jumps a given number of steps in a direction. We specifically studied the impact of this protocol on the first passage time of the 1D infinite lattice random walk. Additionally, we computed the cost~\eqref{cost} of stochastic resetting until the walker hits the target/absorbing boundary for the first time. Our analysis revealed that this dynamic protocol is effective (in contrary to traditional resetting) in finding the target in the limit of large resetting probability $p_r$. The protocol allows to optimize the cost of resetting, and this optimization depends on the nature of cost function. 

This study opens several research avenues in the field of stochastic resetting. It would be interesting to extend the analysis for other cases when the space is either continuous or discrete and time is continuous, including extension to higher dimensions. An interesting question for future investigations (but beyond the scope of this paper) would be to compare results of \dsr\ with that of the periodic dynamical stochastic resetting, where the counter, $\sigma$, [Eq.~\eqref{eqn:fdrp}] resets with probability 1 whenever $|\sigma|=N_r$ and the walker resets to the resetting location with probability $p_r$.

\begin{acknowledgements}
This study was supported by the Special Research Fund (BOF) of Hasselt University (BOF number: BOF24KV10).
This research was enabled in part by support provided by BC DRI Group and the Digital Research Alliance of Canada (\url{www.alliancecan.ca}).
\end{acknowledgements}
\renewcommand{\thefigure}{\arabic{figure}}
\renewcommand{\appendixname}{Appendix}
\appendix
\setcounter{figure}{0}
\renewcommand{\thefigure}{\thesection\arabic{figure}}

\section{Probability distribution function for $N_r=1$}
\label{sec:pdf}
In this section, we compute the probability distribution function of a random walker  to be at position $x$ in time $n$ starting from initial position $x=\xinit$ undergoing the tradition stochastic resetting mechanism  with $N_r = 1$ (in the absence of absorbing boundary). In this case, after every spatial jump a reset check is made. Hence, with probability $0\leq p_r\leq 1$, the random walker resets to position $x=\xinit$; otherwise, the random walker hops to left or right according to the underlying dynamical model. To compute the probability distribution $P(x,n|\xinit)$, we compute the contribution from different trajectories that depend on the number of resetting events
\begin{align}
    P(x,n|\xinit)=\sum_{m=0}^{\infty}\psi_m(x,n|\xinit)\ ,
\end{align}
with $\psi_m(x,n|\xinit)$ the contribution from all trajectories for which exactly $m$ resets occur.

For trajectories without a single reset, the contribution to the probability distribution is
\begin{align}\label{psi0}
    \psi_0(x,n|\xinit) &= (1-p_r)^n P_{\rm NR}(x,n|\xinit)\ ,
\end{align}
where $(1-p_r)^n$ be the probability of not a single resetting event up to time $n$, and during that duration the walker freely propagates with the probability distribution $P_{\rm NR}(x,n|\xinit)$. For one resetting event, the contribution is 
\begin{align}\label{psi1}
        \psi_1(x,n|\xinit) &=\sum_{m=1}^n~\underbrace{(1 - p_r)^{m-1}p_r }_{I_1}~\underbrace{(1-p_r)^{n-m} P_{\rm NR}(x,n-m|\xinit)}_{I_2}\ ,
\end{align}
where $I_1$ is the term corresponding to the random walker not resetting the first $m-1$ times, and then, system resets to $\xinit$ at $m$-th time. $I_2$ corresponds to those trajectories which have not reset in the remaining $n-m$ time. Therefore, we write Eq.~\eqref{psi1} using Eq.~\eqref{psi0}:
\begin{align}\label{psi1-1}
        \psi_1(x,n|\xinit) &=p_r\sum_{m=1}^n (1-p_r)^{m-1}~\psi_0(x,n-m|\xinit) \ .
\end{align}
Similarly, we can write the contribution to the probability distribution for $\mathcal{N}_r$ resets by generalizing Eq~\eqref{psi1-1}:
\begin{align}\label{psi1-2}
        \psi_{\mathcal{N}_r}(x,n|\xinit) &=p_r\sum_{m=1}^n (1-p_r)^{m-1}~\psi_{\mathcal{N}_r-1}(x,n-m|\xinit) \ .
\end{align}
The above equation corresponds to the first-renewal mechanism. To compute the probability distribution $P(x,n|\xinit)$, we $z$-transform Eq.~\eqref{psi1-2}, and it gives
\begin{align}
    \tilde{\psi}_{\mathcal{N}_r}(x,z|\xinit) =&\dfrac{p_r}{1-p_r} \sum_{n=0}^{\infty}z^n\bigg[\sum_{m=0}^n (1 - p_r)^{m}- \delta_{m,0} \bigg]\psi_{\mathcal{N}_r-1}(x,n-m|\xinit)\ ,\\
    =&\dfrac{p_r}{1-p_r}\bigg[  \sum_{m=0}^{\infty}\sum_{n=m}^\infty z^n(1 - p_r)^{m} \psi_{\mathcal{N}_r-1}(x,n-m|\xinit) - \tilde{\psi}_{\mathcal{N}_r-1}(x,z|\xinit)\bigg]\ ,\\
    =& \dfrac{zp_r}{1-z(1-p_r)}\tilde{\psi}_{\mathcal{N}_r-1}(x,z|\xinit)\ . \label{recur}
\end{align}
The above equation~\eqref{recur} is a recursive relation, which can be simplified as
\begin{align}
       \tilde{\psi}_{\mathcal{N}_r}(x,z|\xinit) =& \bigg[\dfrac{zp_r}{1-z(1-p_r)}\bigg]^{\mathcal{N}_r}\tilde{\psi}_{0}(x,z|\xinit)\ .
\end{align}
Then, the full propagator in $z$-space can be written by summing over $\mathcal{N}_r$ from $0$ to $\infty$, and this gives 
\begin{align}
    \tilde{P}(x,z|\xinit) &= \dfrac{1 - z(1-p_r)}{1-z} \psi_0(x,z|\xinit)\ ,\\
    &=\dfrac{1 - z(1-p_r)}{1-z} \tilde{P}_{\rm NR}(x,z(1-p_r)|\xinit)\ . \label{pz-sol}
\end{align}
The above equation connects the propagator of the random walk under resetting $\tilde{P}(x,z|\xinit)$ with that of in the absence of resetting $\tilde{P}_{\rm NR}(x,z|\xinit)$ [see Eq.~\eqref{SS-AC}].
Then, the stationary state distribution is obtained as
\begin{align}
    P^{\rm ss}(x) \equiv \lim_{z\to 1} [(1-z)\tilde{P}(x,z|\xinit) ] =  p_r\tilde{P}_{\rm NR}(x,1-p_r|\xinit)\ ,\label{pss-x}
\end{align}
where 
\begin{align}\label{SS-AC}
    \tilde{P}_{\rm NR}(x,z|\xinit)\equiv \dfrac{1}{\sqrt{1-4pqz^2}}\bigg[\dfrac{1-\sqrt{1-4pqz^2}}{2zq}\bigg]^{|x - \xinit|}
\end{align}
the $z$-transformed probability distribution function of the random walker starting from $\xinit$ and reaching at $x$ in the absence of both resetting mechanism and absorbing boundary $\tilde{P}_{\rm NR}(x,z|\xinit)$ is (see Eq.~(1.3.11) in~\cite{Redner}). See Ref.~\cite{D_Das_2022} for a derivation of the stationary distribution of a random walk under stochastic resetting using the last renewal method and a different form of reset-free distribution $\tilde{P}_{\rm NR}(x,z|\xinit)$~\cite{non-reset-distribution} . 

\section{Moment generating function of first passage time and cost for $N_r=1$}
In the following, we compute the moment generating function of the cost, $C_{\beta}$~\eqref{cost}:
\begin{align}
    \Phi(k,n_{\rm FP}|x_0) \equiv \big\langle e^{ikC_\beta}\big\rangle = \sum_{\mathcal{N}_r=0}^{\infty}\Psi_{\mathcal{N}_r}(k,n_{\rm FP}|x_0)\ ,
\end{align}
where $\mathcal{N}_r$ is the number of resets until the random walker (performing traditional stochastic resetting, i.e., $N_r=1$) hits the target for the first time, and the angular brackets indicate the averaging over ensemble of these trajectories.
We perform the calculation by counting the contributions of each random walker's trajectory starting from $x=\xinit$, stochastically resetting to $x=\xinit$, and hitting the target at $x=\xtarget$ for the first time at $n_{\rm FP}$. We calculate the contributions of such trajectories for a given number of resets. (For convenience, henceforth we drop the subscript `FP' from $n_{\rm FP}$.)
\begin{itemize}
    \item The contribution of trajectories of not having a single reset event (i.e., the random walker hits the target before the first resetting event) to the cost's~\eqref{cost} characteristic function is
    \begin{align}\label{Phi0}
        {\Psi}_0(k, n|\xinit) &= (1-p_r)^n F_{\rm NR}(n|\xinit)\ ,
    \end{align}
    where the first and second terms, respectively, on the right-hand side (rhs) are the probability of not resetting up to the first passage time $n$ and the first passage distribution of a reset-free random walker starting from $x=\xinit$ and hitting the target $x=\xtarget$ for the first time. Notice that for this case, the cost $C_\beta$ is zero.
    \item For one resetting event, we have
    \begin{align}
            {\Psi}_1(k,n|\xinit) &=p_r~\sum_{m=1}^n (1 - p_r)^{m-1}~{\phi}_{\rm NR}(k, m-1|\xinit)~(1 - p_r)^{n-m}~F_{\rm NR}(n - m|\xinit)\ , \label{phi-10}\\
             &=p_r~\sum_{m=1}^n (1 - p_r)^{m-1}~{\phi}_{\rm NR}(k, m-1|\xinit)~{\Psi}_0(k,n-m|\xinit)\ , \label{Phi-1}
    \end{align}
    where we used $F_{\rm NR}(0|\xinit)=0$ in Eq.~\eqref{phi-10}, Eq.~\eqref{Phi0} in  Eq.~\eqref{Phi-1}, and defined the reset-free (indicated by subscript `NR') moment generating function of the cost:
    \begin{align}\label{mgf-r-0}
        {\phi}_{\rm NR}(k, m|\xinit) \equiv \sum_{x=-\infty}^{\xtarget}e^{i k |x-\xinit|^\beta} P_{\rm NR}^{\rm abs}(x,n|\xinit)\ ,
    \end{align}
    for the absorbing boundary at $x=\xtarget$ and the reset-free random walker's probability distribution, $P_{\rm NR}^{\rm abs}(x,n|\xinit)$, to be at $x$ starting from $x=\xinit$ in the presence of absorbing boundary at $x=\xtarget$. Notice that for $k=0$, rhs of Eq.~\eqref{mgf-r-0} is the reset-free random walker's survival probability, $S_{\rm NR}(n|\xinit)$.
    \item Since the process renews after each resetting event, we generalize the contribution for $\mathcal{N}_R$ resets by looking at Eq.~\eqref{Phi-1}:
    \begin{align}\label{Phir}
        {\Psi}_{\mathcal{N}_r}(k,n|\xinit)=p_r~\sum_{m=1}^n (1 - p_r)^{m-1}~{\phi}_{\rm NR}(k, m-1|\xinit)~{\Psi}_{\mathcal{N}_r-1}(k,n-m|\xinit)\ .
\end{align}
\end{itemize}
In order to proceed, we $z$-transform the above equation~\eqref{Phir}:
\begin{align}
      \tilde{{\Psi}}_{\mathcal{N}_r}(k,z|\xinit) &=p_r \sum_{n=0}^\infty~z^n~\sum_{m=0}^{n-1}~(1-p_r)^m {\phi}_{\rm NR}(k, m|\xinit)~{\Psi}_{\mathcal{N}_r-1}(k,n-m-1|\xinit) \\
       &=z p_r \sum_{m=0}^\infty~z^m~(1-p_r)^{m} {\phi}_{\rm NR}(k, m|\xinit)~\sum_{n=1}^{\infty} z^{n-1}{\Psi}_{\mathcal{N}_r-1}(k,n-1|\xinit) \\
       & = zp_r~\tilde{{\phi}}_{\rm NR}(k, z(1-p_r)|\xinit)~\tilde{{\Psi}}_{\mathcal{N}_r-1}(k,z|\xinit) \\
       & = [zp_r~\tilde{{\phi}}_{\rm NR}(k, z(1-p_r)|\xinit)]^{\mathcal{N}_r}~\tilde{{\Psi}}_{0}(k,z|\xinit)\ .
\end{align}
Summing over all reset numbers $\mathcal{N}_r$ from $0$ to $\infty$, we get
\begin{align}\label{mgf-cb}
\tilde{{\Phi}}(k,z|\xinit)& = \dfrac{\tilde{F}_{\rm NR}(z(1-p_r)|\xinit)}{1-zp_r \tilde{{\phi}}_{\rm NR}(k, z(1-p_r)|\xinit) }\ ,
\end{align}
where we used the $z$-transformed $\tilde{{\Psi}}_0(k, z|\xinit)$ [see Eq.~\eqref{Phi0}].
\subsection{Moment generating function for first passage time}
\label{sec:mgf-fpt}
For Fourier variable $k=0$, the above Eq.~\eqref{mgf-cb} gives relation between the moment generating function of the first passage time in the presence of resetting with that of in the absence of resetting:
\begin{align}\label{fpt-st}
\tilde{F}(z|\xinit)& = \dfrac{\tilde{F}_{\rm NR}(z(1-p_r)|\xinit)}{1-zp_r \tilde{S}_{\rm NR}(z(1-p_r)|\xinit) }\ .
\end{align}
We can write the above relation~\eqref{fpt-st} in terms of survival probability by noticing the relation between the first passage distribution at time $n$ and the survival probability:
\begin{align}
    F(n|\xinit) = S(n-1|\xinit) - S(n|\xinit)\ .
\end{align}
The $z$-transform of this gives 
\begin{align}\label{FS-rel}
    \tilde{F}(z|\xinit) = (z-1)\tilde{S}(z|\xinit) + 1\ ,
\end{align}
which ensures normalization for $z=1$, as expected. 

Substituting Eq.~\eqref{FS-rel} on the lhs of Eq.~\eqref{fpt-st} and rearranging terms, we get the $z$-transformed survival probability in the presence of resetting:
\begin{align}\label{fpt-st-2}
\tilde{S}(z|\xinit)& = \dfrac{1}{z-1}\dfrac{\tilde{F}_{\rm NR}(z(1-p_r)) + z p_r \tilde{S}_{\rm NR}(z(1-p_r)) - 1}{1-zp_r \tilde{S}_{\rm NR}(z(1-p_r)) }\ .
\end{align}

Notice the relation~\eqref{FS-rel} also holds for reset-free dynamics. Therefore, replacing $z\to z(1-p_r)$, we get 
\begin{align}\label{FS-rel-2}
    \tilde{F}_{\rm NR}(z(1-p_r)|\xinit) = (z(1-p_r)-1)\tilde{S}_{\rm NR}(z(1-p_r)|\xinit) + 1\ .
\end{align}

Substituting Eq.~\eqref{FS-rel-2} on the rhs of Eq.~\eqref{fpt-st-2}, we get  
\begin{align}\label{fpt-fin}
    \tilde{S}(z|\xinit) = \dfrac{\tilde{S}_{\rm NR}(z(1-p_r))}{1-zp_r \tilde{S}_{\rm NR}(z(1-p_r)) }\ .
\end{align}
This equation (already derived in Ref.~\cite{kusmierz2014first, D_Das_2022}) provides the connection between survival probabilities of reset and reset-free dynamics. In the following, we discuss the moments of the first passage time.

Differentiating both sides of Eq.~\eqref{FS-rel} with respect to $z$, we get 
\begin{align}\label{FS-rel-3}
    \tilde{F}'(z|\xinit) = (z-1)\tilde{S}'(z|\xinit) + \tilde{S}(z|\xinit) \ ,
\end{align}
which for $z=1$ becomes
\begin{align}\label{mfpt-def}
    \tilde{F}'(1) = \tilde{S}(1)\ .
\end{align}
Here lhs is the mean first passage time $\langle n_{\rm FP}\rangle$; therefore, $\langle n_{\rm FP}\rangle$ can be evaluated from the survival probability~\eqref{fpt-fin} by substituting $z=1$ on both sides:
\begin{align}\label{mfpt-1}
    \langle n_{\rm FP} \rangle = \dfrac{\tilde{S}_{\rm NR}(1-p_r)}{1-p_r \tilde{S}_{\rm NR}(1-p_r)}\ .
\end{align}

Similarly, taking one more derivative of Eq.~\eqref{FS-rel-3} with respect to $z$ and then setting $z=1$, we get
\begin{align}\label{f2p}
    \tilde{F}''(1) = 2 \tilde{S}'(1)\ ,
\end{align}
where the lhs is $\langle n_{\rm FP}^2 \rangle - \langle n_{\rm FP}\rangle$. Then, using the mean first passage time~\eqref{mfpt-1}, we can compute the second moment of first passage time. 

The calculation shown in this section are valid for any dimensions. In the following, we specialize for the case of one dimensional random walker. To compute the mean first passage time~\eqref{mfpt-1}, we recognize the relation between the survival probability and the probability distribution function of the random walker~(see Eq.~(1.2.2) in~\cite{Redner}):
\begin{align}\label{SP-AC}
    \tilde{S}_{\rm NR}(z|\xinit) = \dfrac{1}{1-z}\bigg[1 - \dfrac{\tilde{P}_{\rm NR}(x,z|\xinit)}{\tilde{P}_{\rm NR}(x,z|x)}\bigg]\ .
\end{align}
where $\tilde{P}_{\rm NR}(x,z|x)$ is given in Eq~\eqref{SS-AC}.
We emphasize that the relation~\eqref{SP-AC} holds even for the case of resetting dynamics. 

Substituting Eq.~\eqref{SP-AC} by setting $z=1-p_r$ in Eq.~\eqref{mfpt-1} gives the mean first passage time:
\begin{align}\label{mfpt-exp}
   \langle n_{\rm FP} \rangle = \dfrac{2^{| \xtarget-\xinit| }T_1^{-1}-1}{p_r}\ ,
\end{align}
for 
\begin{align}
    T_1 & \equiv \bigg[\dfrac{1 - T_2}{q(1-p_r)}\bigg]^{|\xtarget-\xinit|}\ ,\\
    T_2 &\equiv \sqrt{1-4 p q (1-p_r)^2}\ .
\end{align}
Here, $\xtarget$ and $\xinit$, respectively, target location and initial position of the random walker. Similarly, we can compute the variance of the first passage time using Eqs.~\eqref{f2p} and~\eqref{mfpt-exp}:
\begin{align}
    \langle [n_{\rm FP} - \langle n_{\rm FP} \rangle]^2 \rangle &= \dfrac{T_2(4^{| \xtarget-\xinit| }+p_rT_1^2)-2^{| \xtarget-\xinit|}T_1[T_2 + p_r (T_2 + | \xtarget-\xinit|)]}{p_r^2  T_1^2 T_2}~+ \nonumber\\ &+\langle n_{\rm FP} \rangle - \langle n_{\rm FP} \rangle^2 \\
    & = \dfrac{4^{| \xtarget-\xinit| }T_2 - 2^{|\xtarget-\xinit|}T_1T_2 - 2^{|\xtarget-\xinit|}p_r T_1 |\xtarget - \xinit| }{p_r^2  T_1^2 T_2}- \langle n_{\rm FP} \rangle^2\ .\label{var-exp}
\end{align}

\subsection{Moment generating function of cost}
\label{sec:mgf-cost}
Substituting $z=1$ in Eq.~\eqref{mgf-cb} gives the moment generating function of the cost~\eqref{cost} averaged over all first passage times:
\begin{align}\label{mgf-cost}
\tilde{{\Phi}}(k,1|\xinit)& = \dfrac{\tilde{F}_{\rm NR}(1-p_r|\xinit)}{1-p_r \tilde{{\phi}}_{\rm NR}(k, 1-p_r|\xinit) }\ .
\end{align}

Differentiating with respect to $ik$ and setting $k=0$, we find the mean cost of resetting:
\begin{align}\label{cbeta-aexp}
    \langle C_{\beta} \rangle &= p_r\dfrac{\tilde{F}_{\rm NR}(1-p_r|\xinit)}{[1-p_r {\tilde{S}}_{\rm NR}(1-p_r|\xinit)]^2} \sum_{x=-\infty}^{x=\xtarget} |x|^\beta \tilde{P}^{\rm abs}_{\rm NR}(x,1-p_r|\xinit)\\
    &=p_r \langle n_{\rm FP}\rangle \dfrac{\sum_{x=-\infty}^{x=\xtarget} |x|^\beta \tilde{P}^{\rm abs}_0(x,1-p_r|\xinit)}{{S}_{\rm NR}(1-p_r|\xinit)}\ ,\label{cost-beta}
\end{align}
where $\tilde{P}^{\rm abs}(x,z|\xinit)$ is the $z$-transformed probability distribution of a random walker to be at $x$ starting from $x=\xinit$ in the presence of absorbing boundary at $x=\xtarget$. Clearly, for $\beta=0$, the mean cost (which is the average number of resets)
\begin{align}\label{c0-formula}
    \langle C_{0} \rangle = p_r \langle n_{\rm FP}\rangle\ ,
\end{align}
as expected. 

For a symmetric 1D random walk ($p=q$), $\tilde{P}_{\rm NR}^{\rm abs}(x,z)$ can be computed using the method of images:
\begin{align}
    \tilde{P}_{\rm NR}^{\rm abs}(x,z|\xinit) \equiv \tilde{P}_{\rm NR}(x,z|\xinit) - \tilde{P}_{\rm NR}(2\xtarget - x,z|\xinit)\ , \label{image}
\end{align}
for $\xinit < \xtarget$. Using this relation~\eqref{image} we can compute the mean cost~\eqref{cost-beta}.

\begin{figure} % figure B.1
    \centering
    \includegraphics[width = \textwidth]{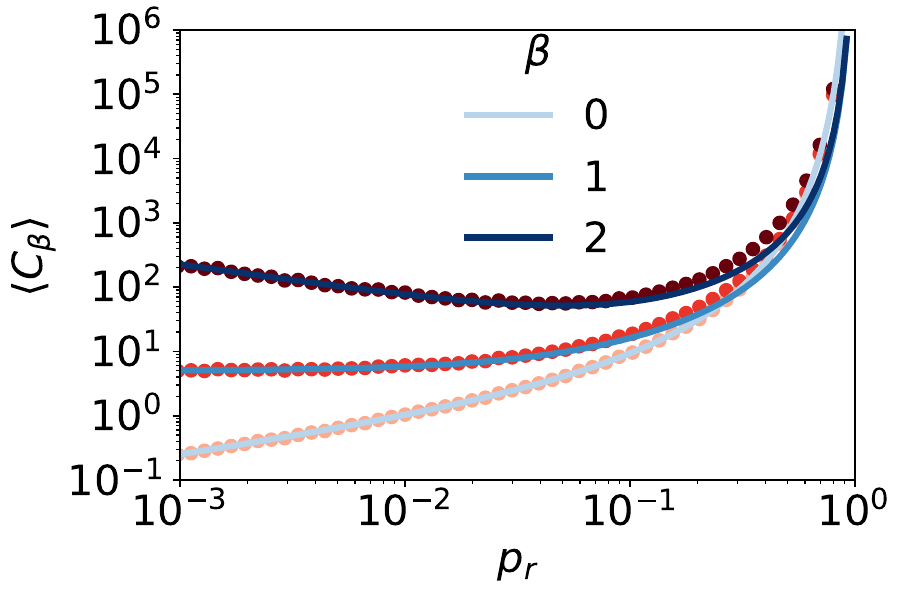}
    \caption{Unbiased random walk ($p=q=1/2$). Mean cost, $\langle C_{\beta}\rangle$~\eqref{cost}, as a function of resetting probability $p_r$. Symbols (red): Numerical simulation data. Curves (blue): Analytical prediction~\eqref{cost-beta} for $N_r = 1$. Here, the target is at $\xtarget=5$. The color intensity increases with the exponent $\beta$.}
    \label{fig:analytical-sim-cost}
\end{figure}

Figure~\ref{fig:analytical-sim-cost} displays the comparison of analytical result~\eqref{cost-beta} mean cost $\langle C_\beta\rangle$ with that of numerical simulation data for $N_r = 1$ and for three $\beta$ values. We find small disagreement for large $p_r$ value; this might be because of numerical evaluatation of the summation in Eq.~\eqref{cost-beta}. 

\section{Mean first passage time: $p_r\to 1$}
\setcounter{figure}{0}
\label{sec:mfpt-pr-to-1}
Figure~\ref{fig:mfpt-pr-1} discuss the mean first passage time for the random walker in the presence of dynamic resetting for $p_r = 1$. 
For $N_r =1$ the MFPT diverges for for all $\xtarget \geq 1$, whereas MFPT diverges for $\xtarget\geq 2$ for $N_r = 2$. For other $N_r$, MFPT is finite. 
MFPT is optimized with respect to $N_r$, and it has minimum value for $N_r^* = \xtarget + 1$.

\begin{figure} % figure C.1
    \centering
    \includegraphics[width = \textwidth]{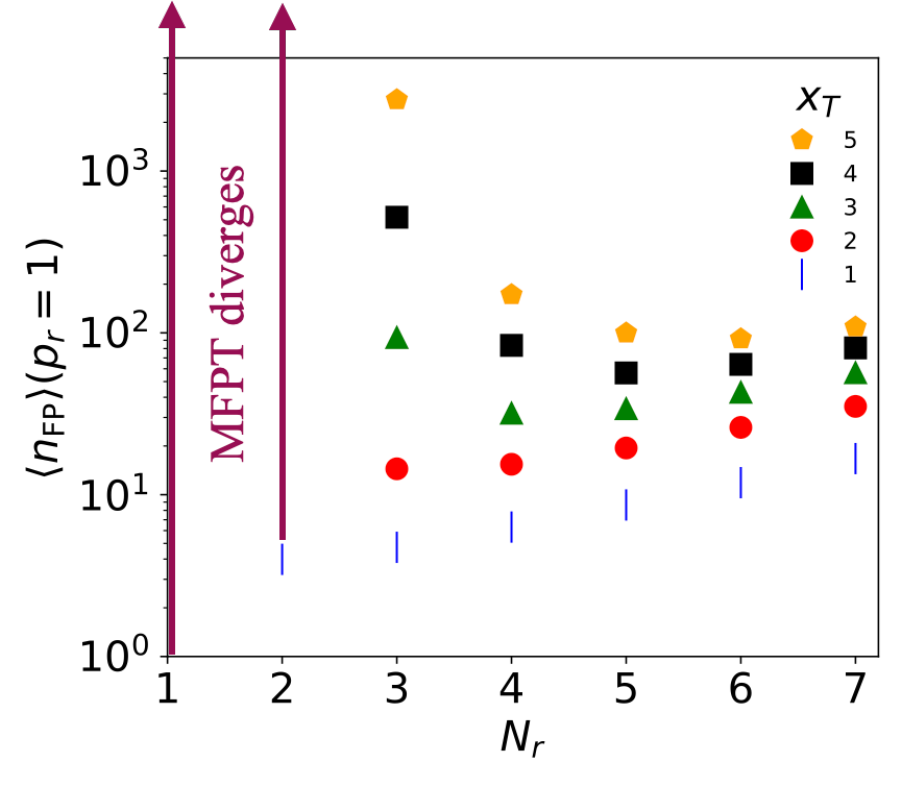}
    \caption{Unbiased random walk ($p=q=1/2$). Mean first passage time (MFPT) at resetting probability $p_r = 1$, as a function of threshold distance $N_r$ for different target positions. For $N_r=1$ the RW remains indefinitely at the reset position and the MFPT diverges as soon as $\xtarget \neq \xreset$. For $N_r=2$ the MPFT is finite for $\xtarget=1$ and diverges otherwise. Symbols: Numerical simulation data.}
    \label{fig:mfpt-pr-1}
\end{figure}

\section{Mean first passage time: Dynamic {\it vs.}~window  stochastic resetting}
\setcounter{figure}{0}
\label{sec:window-resetting}
In this section, we compare the MFPT obtained from two different resetting schemes: 1) dynamic stochastic resetting (\dsr), and 2) window stochastic resetting (wSR). In contrast to the \dsr, wSR involves stochastically resetting the random walker to its initial location $\xinit$ with probability $p_r$ as soon as it hits either of the boundary of the window: 
\begin{align}\label{eq:wsr}
    x_n = \begin{cases}
        x_{n-1} + 1&\qquad {\rm with~probability~}p\\
        x_{n-1} - 1&\qquad {\rm with~probability~}1-p\\
        \xinit &\qquad {\rm with~probability~}~p_r~{\rm if}~|x_n|\geq N_r\ . 
    \end{cases}
\end{align}
Notice that wSR for a continuous space and continuous time is studied in Ref.~\cite{evans2011optimal}.
We remind that $N_r$ in \dsr\ is the threshold of the total number of steps taken by the random walker in one direction. All resetting protocols, ie. \dsr, wSR and \tsr, are identical for $N_r = 1$.

\begin{figure}
    \centering
    \includegraphics[width = \textwidth]{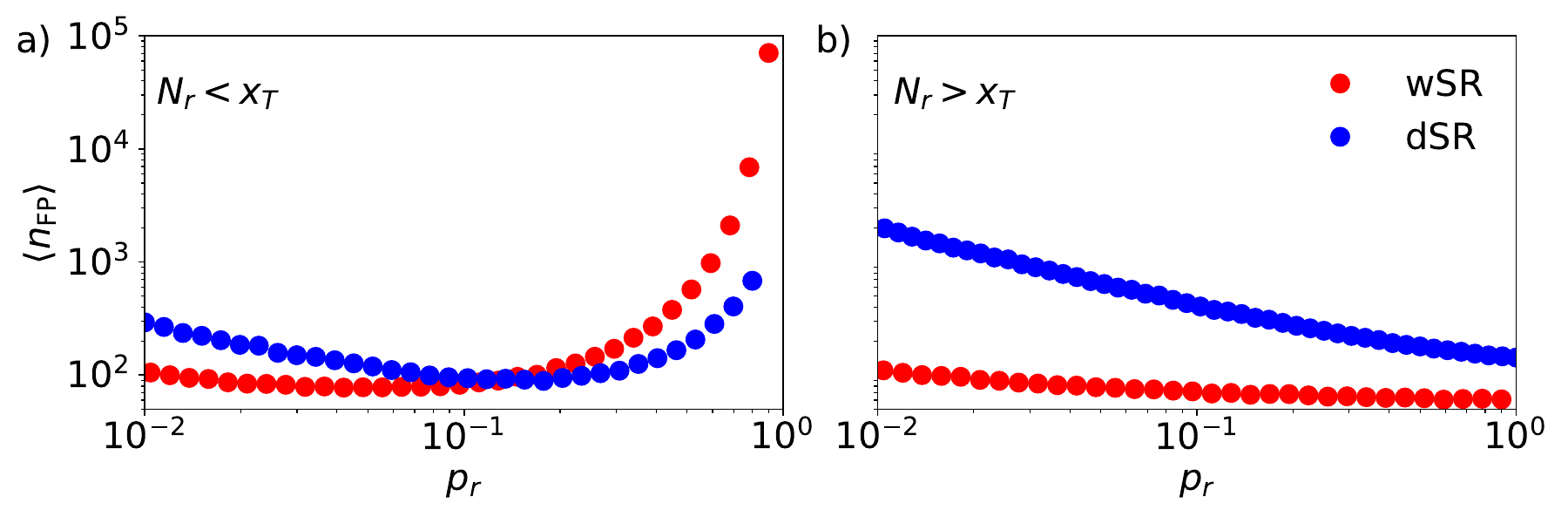}
    \caption{Unbiased Random walk ($p=q=1/2$). Blue: dynamic stochastic resetting (\dsr)~\eqref{eqn:fdrp}. Red: Window stochastic resetting (wSR)~\eqref{eq:wsr}. Threshold distance a) $N_r = 3$. b) $N_r = 7$. Target distance from resetting location $\xtarget = 5$.}
    \label{fig:WSR-DSR}
\end{figure}

Figure~\ref{fig:WSR-DSR} shows the comparison of the MFPT of \dsr\ with that of wSR for two different scenarios: (1) $N_r < \xtarget$, and (2) $N_r > \xtarget$. \dsr\ has lower MFPT in scenario (1) for large $p_r$, whereas for scenario (2) wSR is better than \dsr\ for all $p_r$. 

\section{Cost of resetting for $N_r\geq \xtarget$}
\setcounter{figure}{0}
\label{sec:cost-nr-gr-xt}
Figure~\ref{fig:cost-nr-gr-xt} discusses mean cost $\langle C_\beta \rangle$ per mean first passage time as a function of resetting probability $p_r$, for $N_r\geq \xtarget$. This ratio decreases as $N_r$ increases for $\beta\leq 1$, and for $\beta=2$, this is insensitive to $N_r$ (as also seen in Fig.~\ref{fig:cost0}).   

\begin{figure}
    \centering
    \includegraphics[width = \textwidth]{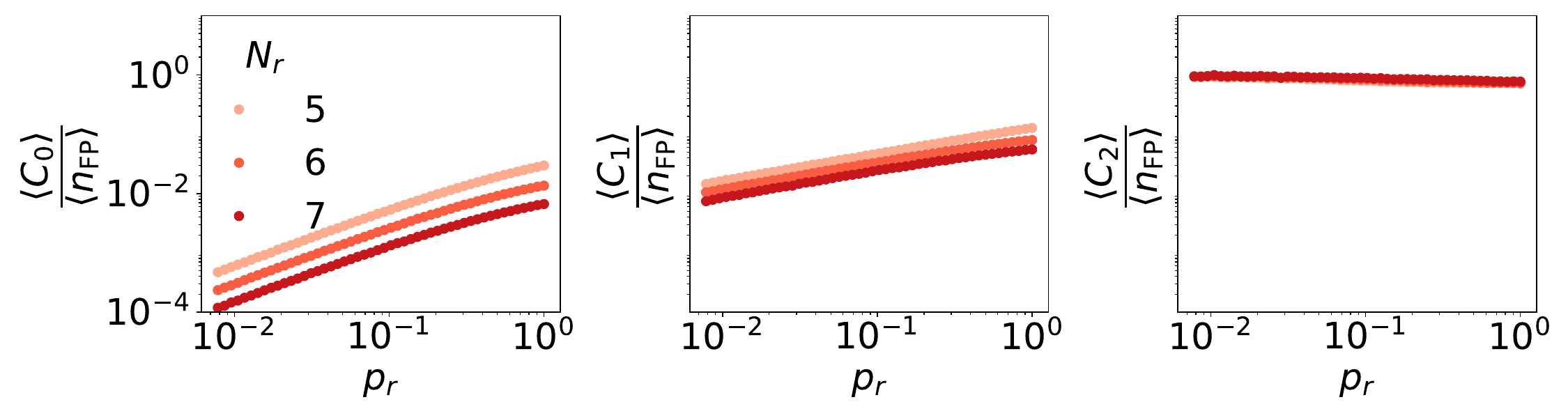}
    \caption{Unbiased random walk ($p=q=1/2$). Mean cost $\langle C_{\beta}\rangle$~\eqref{cost} per mean first passage time as a function of resetting probability $p_r$. Symbols (red): Numerical simulation data. Here, the target is at $\xtarget=5$. The color intensity increases with the exponent $N_r$.}
    \label{fig:cost-nr-gr-xt}
\end{figure}

\section{Method of simulations}
\label{methods}
In the following, we present the numerical simulation method to compute the first passage time and the cost $C_\beta$ for 1D case. For convenience, we place the absorbing boundary $\xtarget$ on the positive side of the origin, and reset the particle to the origin. We follow the dynamical rules as described in Eq.~\eqref{eqn:fdrp}. To proceed, we initialize the position of the walker location ($x=\xinit$), the counter ($\sigma = 0$), first passage time ($n_{\rm FP}=0$), and the cost ($C_\beta = 0$). Then, for each fixed $\xtarget$ and $N_r$ (threshold distance), we generate the following trajectory: 
\begin{enumerate}
    \item \label{step-1} If $x\geq \xtarget$ the simulation stops [we go to step~\eqref{fin-step}]; otherwise, it continues as below.  
    \item With probability $p$ / $1-p$, the random walker jumps to the right/left, and increase/decrease $\sigma$ by one unit if it has a positive / negative value in the previous step; otherwise, we set it to $+1/-1$.
    \item Next if $|\sigma| \geq N_r$, we reset the random walker to the resetting location with probability $p_r$ and update the cost by an amount of $|x-\xreset|^\beta$, where $x$ and $\xreset$, respectively, are random walker's positions just before and after the reset. 
    \item We update the position $x$ and advance the time by one unit.
    \item We go back to step~\eqref{step-1}. This process iterates until the random walker hits the absorbing boundary (i.e., $x \to \xtarget$). 
    \item \label{fin-step} We record the first passage time $n_{\rm FP}$ and cost $C_\beta$.   
\end{enumerate}
We repeat the above algorithm for $\mathcal{N}_R$ realizations to compute the statistics of first passage time and cost $C_\beta$. 

\section*{References}
% \bibliographystyle{iopart-num}
% \bibliography{ref.bib,refs_anm.bib}

\providecommand{\newblock}{}

\end{document}